\documentclass[sigconf]{acmart}

\usepackage{multirow}
\usepackage{tabularx,booktabs}
\usepackage{amsmath}
\usepackage{amsfonts}
\usepackage{amsbsy}
\usepackage{enumitem}

\usepackage{xcolor}
\usepackage{colortbl}
\definecolor{graylite}{gray}{.89}
\newcolumntype{g}{>{\columncolor{graylite}}c}

\AtBeginDocument{%
  \providecommand\BibTeX{{%
    \normalfont B\kern-0.5em{\scshape i\kern-0.25em b}\kern-0.8em\TeX}}}

\setcopyright{acmcopyright}

\copyrightyear{2022}
\acmYear{2022}
\setcopyright{acmcopyright}\acmConference[CIKM '22]{Proceedings of the 31st ACM International Conference on Information and Knowledge Management}{October 17--21, 2022}{Atlanta, GA, USA}
\acmBooktitle{Proceedings of the 31st ACM International Conference on Information and Knowledge Management (CIKM '22), October 17--21, 2022, Atlanta, GA, USA} \acmPrice{15.00}
\acmDOI{10.1145/3511808.3557317}
\acmISBN{978-1-4503-9236-5/22/10}

\begin{document}

\title{Explanation Guided Contrastive Learning for Sequential Recommendation}

\author{Lei Wang, Ee-Peng Lim*}\thanks{*Corresponding Author.}
\affiliation{%
  \institution{Singapore Management University}
  \country{Singapore}
}
\email{{lei.wang.2019@phdcs., eplim@}smu.edu.sg}

\author{Zhiwei Liu}
\affiliation{%
  \institution{Salesforce}
    \country{USA}
}
\email{zhiweiliu@salesforce.com}

\author{Tianxiang Zhao}
\affiliation{%
  \institution{Penn State University}
    \country{USA}
}
\email{tkz5084@psu.edu}

\renewcommand{\shortauthors}{Lei Wang, Ee-Peng Lim, Zhiwei Liu, \& Tianxiang Zhao}

\begin{abstract}
Recently, contrastive learning has been applied to the sequential recommendation task to address data sparsity caused by users with few item interactions and items with few user adoptions.  Nevertheless, the existing contrastive learning-based methods fail to ensure that the positive (or negative) sequence obtained by some random augmentation (or sequence sampling) on a given anchor user sequence remains to be semantically similar (or different). 
When the positive and negative sequences turn out to be false positive and false negative respectively, it may lead to degraded recommendation performance. In this work, we address the above problem by proposing  \textbf{Explanation Guided Augmentations (EGA)} and \textbf{Explanation Guided Contrastive Learning for Sequential Recommendation (EC4SRec)} model framework.  The key idea behind EGA is to utilize explanation method(s) to determine items' importance in a user sequence and derive the positive and negative sequences accordingly.  EC4SRec then combines both self-supervised and supervised contrastive learning over the positive and negative sequences generated by EGA operations to improve sequence representation learning for more accurate recommendation results. 
Extensive experiments on four real-world benchmark datasets demonstrate that EC4SRec outperforms the state-of-the-art sequential recommendation methods and two recent contrastive learning-based sequential recommendation methods, CL4SRec and DuoRec. Our experiments also show that EC4SRec can be easily adapted for different sequence encoder backbones (e.g., GRU4Rec and Caser), and improve their recommendation performance.\footnote{Code is available at \url{https://github.com/demoleiwang/EC4SRec}.} 
\end{abstract}

\ccsdesc[500]{Information systems~Recommender systems}


\keywords{Sequential Recommendation, Contrastive Learning, Explanation}

\maketitle

\section{Introduction}

\textbf{Background.} Recommender systems have played an important role in today's online services~\cite{covington2016deep, harper2015movielens, mcauley2015image} to help users navigate the overwhelming amount of information and discover interesting items. 
Since sequential patterns of user-item interactions change with time, 
researchers thus ~\cite{hidasi2015session, tan2016improved, wu2017recurrent, tang2018personalized, kang2018self} pay much attention to sequential recommendation which focuses on characterizing dynamics in user sequences
to predict next user-item interaction(s).

\begin{figure}[t]
  \centering
  \includegraphics[width=1.0\linewidth]{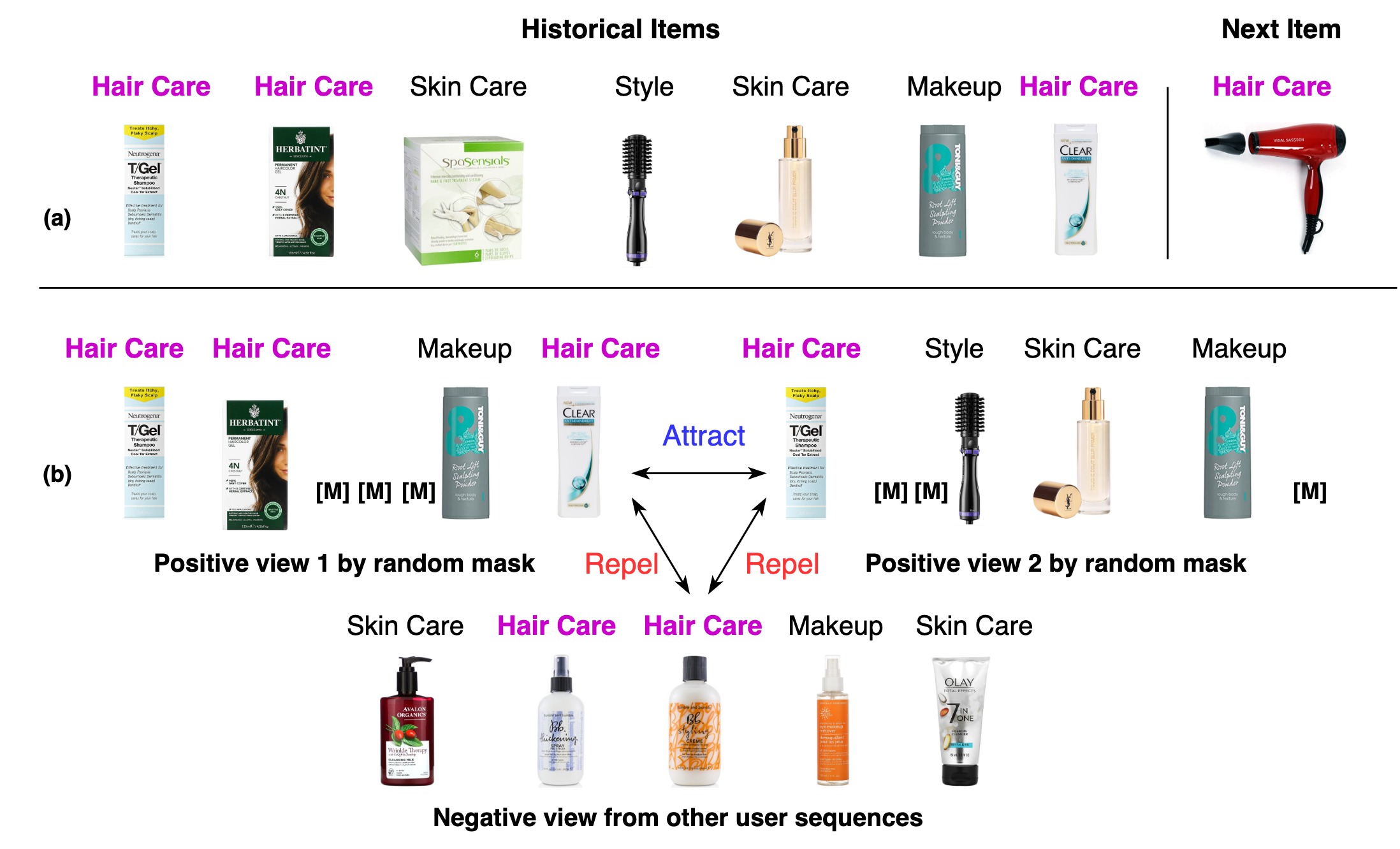}
  \vspace{-10pt}
  \caption{Motivation example: (a) A given user sequence with seven items and a red hair-dryer as the next item; (b) Two positive views generated by random mask operations on the given sequence and a negative view which is the sequence of another user. [M] represents a masked item.}
  \label{fig:example}
\vspace{-10pt}
\end{figure}

For a sequential recommendation method to yield accurate results, it has to learn a high-quality user representation from the user's historical sequence and match the user representation against candidate items. 
Traditional methods model low-order dependencies between users and items via Markov Chain and Matrix Factorization~\cite{rendle2010factorizing, he2016fusing}. Recently, researchers have developed deep learning-based (DL-based) sequential recommendation methods using deep neural networks (such as recurrent neural networks~\cite{hidasi2015session}, convolutional neural networks~\cite{tang2018personalized}, transformer~\cite{kang2018self}, and graph neural networks~\cite{chang2021sequential}) which learn higher-order dependencies to predict the next items. However, data sparsity is still a major challenge due to limited data about users and items in the long tail. The former refers to many users having very short item sequences. The latter refers to many items having very few user interactions. To cope with these challenges, contrastive learning-based (CL-based) sequential recommendation works~\cite{xie2020contrastive, zhou2020s3, liu2021contrastive} incorporate positive and negative views of original user sequences by augmentations and sampling so as to learn more robust user sequence representations, thus more accurately matching candidate items to improve recommendation performance.

\textbf{Motivating Example.}  Figure~\ref{fig:example} shows an example of the contrastive learning approach to sequential recommendation. From a given user sequence shown in Figure~\ref{fig:example}(a), we obtain two positive views of the user sequence using some augmentation operator(s), and select the sequence of another user as a negative view. For the positive views, we randomly mask as few items in the given user sequence as shown in Figure~\ref{fig:example}(b). To learn user sequence representations, contrastive loss(es) is introduced to make the representations of positive views to be close to each other, but far from that of the negative view~\cite{xie2020contrastive, zhou2020s3, liu2021contrastive}. 

Note that even as CL approach has been shown to improve sequential recommendation performance, its user sequence augmentation and sampling methods are performed with randomness (e.g., random crop, random mask, and random reorder) and is thus prone to produce for a given user sequence positive views that look very different and negative views that look quite similar.
As a result, the learned sequence representations are non-ideal reducing the recommendation accuracy.  For example, if we were to know that the red hair dryer is the next item, the hair care items in the original sequence will be considered to be more relevant (or important). The positive view 2 in Figure~\ref{fig:example}(b) which has two hair care items masked however looks quite different from  positive view 1.  Attracting the representations of positive views 1 and 2 to be closer to each other is therefore inappropriate and may degrade the recommendation performance.  By the same reasoning, the negative view may be inappropriately sampled if it shares many hair care items with the two positive views. 

\textbf{Proposed idea.} The above motivating example suggests that we need to carefully choose positive and negative views in order to learn good high-quality user sequence representations.  To begin this research, we thus conduct a small experiment to show that items important to the next-item of the predicted sequence should be treated differently from non-important items for CL-based sequential recommendation to achieve high accuracy. While this result is interesting, it is infeasible to know which items are important in a user sequence as the next-item is not given during model training. 
To determine the elusive ``important items'', we therefore propose \textbf{explanation guided augmentations} (\textbf{EGA}) to infer the important items of a given user sequence using explanation methods and consider item importance in augmentation operations.  This way, better positive views and negative views can be derived for contrastive learning.  We also propose the \textbf{Explanation Guided Contrastive Learning for Sequential Recommendation (EC4SRec)} framework to utilise the positive and negative views for self-supervised and supervised learning of user sequence representations, combined with recommendation loss function.
EGA and EC4SRec are also designed to accommodate different sequential recommendation backbones. 
In other words, they can be readily applied to existing self-supervised and supervised contrastive learning methods to improve their recommendation performance.

\textbf{Our contributions.}
In summary, our contribution is three-fold:
\begin{itemize}[leftmargin=*]
    \item We propose a model-agnostic Explanation Guided Contrastive Learning for Sequential Recommendation (EC4SRec) framework that incorporates explanation methods into user sequence augmentations for generating positive and negative user sequences for both self-supervised and supervised contrastive learning. EC4SRec can be seen as an integration of CL4SRec and DuoRec, with an additional sampling of negative views for  contrastive learning to more effectively separate the representations of positive views from that of the negative views.  To our knowledge, EC4SRec is also the first that utilizes explanation methods to improve sequential recommendation. 
  \item We propose several explanation guided augmentation operations to generate both positive and negative user sequences using importance score derived from explanation methods. With these operations, EC4SRec can effectively utilize augmented positive and negative user sequences in contrastive learning to obtain better sequence representations for recommendation.
  \item 
  We evaluate the proposed augmentation method over two types of contrastive learning frameworks, with three different base sequential recommendation models, on four real-world datasets.
  The experiment results show that EC4SRec significantly outperforms the vanilla CL4SRec and DuoRec, and other state-of-the-art sequential recommendation methods .  We also demonstrate the generalizability of EC4SRec using different sequence encoders and combinations of explanation methods with consistent performance improvement 
  by $4.2\%$ \textasciitilde $23.0\%$.
\end{itemize} 

\section{Related Work}

\subsection{Sequential Recommendation}
Sequential recommendation aims to learn high-quality user and item representations to predict the next item of a given user sequence. Early works focus on modeling low-order transition relationships between items via Markov Chains as item-item features to be used for recommendation~\cite{rendle2010factorizing, he2016fusing, xu2019graph}. With the advances in neural networks, sequential recommendation research turns to using neural networks~\cite{hidasi2015session, li2017neural, tang2018personalized, kang2018self, zhang2019feature,tan2021sparse, zhang2021causerec, liu2021augmenting, wang2020next, li2020time}, such as RNN~\cite{hidasi2015session}, CNN~\cite{tang2018personalized}, Transformer~\cite{kang2018self}, and GNN~\cite{chang2021sequential} to model high-order sequential dependencies hidden in historical user-item interactions. 
GRU4Rec~\cite{hidasi2015session}, for example, incorporates GRU to model sequence-level patterns. This is further improved by replacing GRU by hierarchical RNN ~\cite{quadrana2017personalizing}. 
Caser~\cite{tang2018personalized} on the other hand uses CNN to model high-order item-item relationships. Inspired by the effectiveness of self-attention in NLP communities~\cite{wiegreffe2019attention}, Kang, et al. ~\cite{kang2018self} apply self-attention in sequential recommendation named SASRec. GNN based models~\cite{liu2021contrastive, wu2019session} are also proposed to capture more complex patterns than sequential patterns. 
To improve sequential recommendation by both performance and interpretability, various works~\cite{huang2018improving, yuan2021dual, chen2018sequential} began to determine items contributing to the next-item prediction. 
Explanation methods, such as attention weights~\cite{wiegreffe2019attention}, gradient-based methods~\cite{zeiler2014visualizing, sundararajan2017axiomatic}, and Occlusion~\cite{simonyan2013deep} have been designed to determine features that explain the prediction labels.  In our research, we explore the use of explanation methods to determine specific earlier items in the user sequence that explain the predicted next-item and in turn improve sequential recommendation accuracy under the EC4SRec framework. 

\subsection{Contrastive Learning}
Contrastive learning has recently achieved great successes in various research domains including computer vision~\cite{chen2020simple, he2020momentum, grill2020bootstrap, peng2022crafting}, NLP~\cite{fang2020cert, gao2021simcse}, recommendation~\cite{zhou2020s3, xie2020contrastive, chen2022intent, wu2021self, chen2022intent, zheng2022disentangling, liu2021self, yao2021self, wu2021self, wang2022contrast_signals, lin2022dual}, etc.. It aims to obtain high-quality representations by pulling positive views of the same instance closer while pushing the positive views and their negative views apart in the representation space.
S$^3$Rec~\cite{zhou2020s3} pre-trains sequential recommendation by contrastive learning with four self-supervised tasks defined on historical items and their attributes.
CL4SRec~\cite{xie2020contrastive} combines recommendation loss with contrastive loss of self-supervised tasks to optimize the sequential recommendation model. 
CoSeRec~\cite{liu2021contrastive} introduces two new augmentation operations, insert and replace, to train robust sequence representations.
DuoRec~\cite{qiu2022contrastive} retrieves the positive view of a given user sequence by finding another user's sequence which shares the same next-item in its proposed supervised contrastive learning. In Section~\ref{sec:prelim}, we will further elaborate CL4SRec and DuoRec.
The above contrastive learning-based sequential recommendation methods, nevertheless, suffer the same pitfalls mentioned in our motivating example. In this research, we therefore seek to address these pitfalls by explanation-guided augmentations and explanation-guided contrastive learning framework.

\subsection{Explanation Methods}
While there are several works on explainable recommendation~\cite{gedikli2014should, tintarev2015explaining, zhang2020explainable}, they are designed to explain why items are recommended by algorithms.  In this work, we mainly focus on general explanation methods~\cite{simonyan2013deep, zeiler2014visualizing, sundararajan2017axiomatic} originally designed to determine features that explain the prediction results.  By applying these methods to sequential recommendation methods, we are able to determine historical items in a user sequence that explain the predicted next-item, and assign importance scores to these historical items.  For example, Saliency~\cite{zeiler2014visualizing}, a widely used explanation method, derives an input feature's attribution score by returning the gradient with respect to the input feature. 
Integrated Gradient~\cite{sundararajan2017axiomatic} takes derivatives of the value for the predicted label with respect to the input features. It outperforms Saliency but is less efficient. 
Models with attention mechanism provide attention weights as the relative importance of items. However, attention as explanation is controversial~\cite{jain2019attention, wiegreffe2019attention} since different attention distributions can produce the same model predictions.
Occlusion~\cite{simonyan2013deep} is a perturbation based explanation method which computes input features' attribution scores by the difference between outputs of the original and perturbed input features. 
\section{Preliminaries}
\label{sec:prelim}

\subsection{Problem Formulation}
Suppose that we have a set of users $\mathcal{U}$ and items $\mathcal{V}$. For the sequential recommendation task, each user $u \in \mathcal{U}$ has a sequence of items the user has interacted with in the past. We denote this sequence by $s_u=[v_1^{u},v_2^{u},\dots,v^{u}_{|s_u|}]$ where $v_i^{u} \in \mathcal{V}$ and $|s_u|$ denotes the sequence length. The items in the sequence are ordered by time.
The goal of sequential recommendation is to predict the next item at time step, i.e., $v_*^{u}$, using the observed historical sequence $s_u$. Suppose $P(v|s)$ is a model that returns the probability of $v$ being the next item given a sequence $s$. The sequential recommendation task can be formulated as:
\begin{center}
    $v_*^{u} = \arg \max_{v \in \mathcal{V}} P\Bigl(v_{|s_u| + 1}^{u}=v \mid s_u\Bigr).$
\end{center}

\subsection{Contrastive Learning for Sequential Recommendation}

In this section, we describe a set of basic augmentation operations to determine positive views of a given user sequence.  These augmentation methods have been used in two latest contrastive learning-based methods, CL4SRec~\cite{xie2020contrastive} and DuoRec~\cite{qiu2022contrastive}.\\

\noindent
\textbf{Basic Augmentation Operations.}
There are four basic augmentation operations~\cite{zhou2020s3, xie2020contrastive,qiu2022contrastive} to generate positive views from an original user sequence, $s_u=[v_1^u,v_2^u,\dots,v^u_{|s_u|}]$. 
\begin{itemize}[leftmargin=*]
    \item \textbf{Random Crop ($\mathrm{crop}$):} It randomly selects a continuous sub-sequence from positions $i$ to $i+l_c$ from $s_u$ and removes it. $l_c$ is defined by $l_c=i+\lfloor\mu_c \cdot |s_u|\rfloor$ where $\mu_c$ ($0 < \mu_c \leq 1$) is a hyper-parameter. The cropped sequence is defined by:\\
    $\ \ \ \ \ s_u^{\mathrm{c}}=[v_i^u,v_{i+1}^u,\dots,v^u_{i+l_c}]$.\\
    \item \textbf{Random Mask ($\mathrm{mask}$):} It randomly selects a proportion $\mu_m$ of items from $s_u$ to be masked. Let $g^m(1), g^m(2), \cdots, g^m(n^m_u)$ be the indexes of the items to be masked where $n^m_{u}=\lfloor \mu_m \cdot |s_u| \rfloor$ and $g^m(x) \in [1,|s_u|]$.  An item $v_i$ is replaced with the mask item [m] if selected to be masked. The masked sequence is thus:\\
      $\ \ s_u^{\mathrm{mask}}=[v_1^u,\cdots,$ $v_{g^m(1)-1}^u, \mbox{[m]}, v_{g^m(1)+1}^u, \cdots, 
      v_{g^m(n^m_u)-1}^u, \mbox{[m]},$\\ 
      $\ \ \ \ \ v_{g^m(n^m_u)+1}^u,  \dots,v^u_{|s_u|}].$\\

    \item \textbf{Random Reorder ($\mathrm{rord}$):} It first randomly selects a continuous sub-sequence $[v_i^u,v_{i+1}^u,\dots,v_{i+l_r}^u]$ of length $l_r=\lfloor\mu_r*|s_{u}|\rfloor$ ($0 \leq \mu_r \leq 1$). It then randomly shuffles the items in the sub-sequence.  Suppose the reordered items, sorted by new positions, are $[\tilde{v}^u_{i}, \dots,\tilde{v}^u_{i+l_r}]$.  The reordered sequence is thus:\\
      $s_u^{\mathrm{rord}}= [v^u_1, \cdots, v^u_{i-1},\tilde{v}^u_{i},\tilde{v}^u_{i+1}, \cdots,\tilde{v}^u_{i+l_r}, v^u_{i+l_r+1}, \cdots v^u_{|s_u|}].$\\

    \item \textbf{Random Retrieval ($\mathrm{rtrl}$):} This operation randomly selects another user sequence $s_{u'}$ that shares the same target (or next) item as the input sequence $s_u$, i.e., $v^u_{*}=v^{u'}_{*}$.  The retrieved sequence is thus:
      $s_u^{\mathrm{rtrl}}=s_{u'}, s.t.\ v^u_{*}=v^{u'}_{*}$ 
\end{itemize}

\noindent
\textbf{CL4SRec Method.}
Consider a set of users in a batch $U_B=\{u_1,u_2,\dots,u_{|U_B|}\}$. The loss function of CL4SRec is:
\begin{equation}
\mathcal{L}_{CL4SRec}=\sum_{u \in U_B} \mathcal{L}_{rec}(s_u) + \lambda \mathcal{L}_{cl}(s_u^{a_i},s_u^{a_j}).
\end{equation}
\noindent
where $\mathcal{L}_{rec}(s_u)$ and $\mathcal{L}_{cl}(s_u^{a_i},s_u^{a_j})$ are the recommendation loss and self-supervised contrastive loss respectively. $s_u^{a_i}$ and  $s_u^{a_j}$ are positive views of original user sequence $s_u$ after applying augmentations $a_i$ and $a_j$ respectively. $a_i$ and $a_j$ are sampled from $\{\mathrm{crop}, \mathrm{mask}, \mathrm{rord} \}$. We denote the positive view pairs for the users in the batch $B$ as $S_B=\{s^{a_1}_{u_1}, s^{a_2}_{u_1}, s^{a_1}_{u_2}, s^{a2}_{u_2},$ $\cdots,$ $s^{a_1}_{u_{|B|}}, s^{a_2}_{u_{|B|}} \}$.  Thus, the recommendation loss for the user $u$ can be formulated as:
\noindent
\begin{equation}
\mathcal{L}_{rec}(s_u)=-\log \frac{\exp(sim(h_u,h_{v^u_*}))}{\exp(sim(h_u,h_{v^u_*}))+\sum_{v^- \in V^-} \exp(sim(h_u,h_{v^-}))},
\end{equation}
where $V^- = V-\{v^u_*\}$,
and $h_{v^{-}}$ are the representations of the sequence $s_u$, the next item $v^*_{u}$, and a negative item $v^{-}$ respectively. 
The contrastive loss is:  
\noindent
\begin{equation}
\begin{split}
\mathcal{L}_{cl}&(s_u^{a_i}, s_u^{a_j})=
 -\log \frac{\exp(sim(h_u^{a_i}, h_u^{a_j}))}{\exp(sim(h_u^{a_i}, h_u^{a_j}))+\sum_{s^- \in S_u^-} \exp(sim(h_u^{a_i}, h^-))},
\end{split}
\end{equation}

\noindent
where $h_u^{a_i}$ and $h_u^{a_j}$ are the representations of $s_u$ after augmentations $a_i$ and $a_j$ respectively. $S_u^-$ denotes a set of negative sequences defined by $S_u^-=S_B - \{s^{a1}_u, s^{a2}_u\}$. $s^-$and $h^-$ denote a sequence that does not belong to $u$ in the batch $B$ and its representation respectively.\\

\textbf{DuoRec Method.}
Given a user sequence $s_u$, we randomly sample a \textit{retrieved-positive view} from other users' sequences that share the same next item $v_*^u$. 
We denote all user sequences and their corresponding retrieved-positive views by $S=\{s_{u_1}, s^\mathrm{rtrl}_{u_1}, s_{u_2}, s^\mathrm{rtrl}_{u_2},$ $\cdots,$ $s_{u_{|B|}}, s^\mathrm{rtrl}_{u_{|U|}} \}$.  
In DuoRec, the representations of each user sequence $s_{u}$ and its retrieved-positive view $s^\mathrm{rtrl}_{u}$ are learned to be close to each other but far from other user sequences and their retrieved-positive views denoted by $S_{u}^-= S - \{s_{u},s^\mathrm{rtrl}_{u}\}$.   

The loss function of DuoRec consists of both recommendation loss and supervised contrastive loss functions:
\begin{equation}
\begin{split}
\mathcal{L}_{DuoRec}=\sum_{u \in U_B} \mathcal{L}_{rec}(s_u) + \lambda \mathcal{L}_{sl}(s_u)
\end{split}
\end{equation}
\begin{equation*}
\begin{split}
\mathcal{L}_{sl}&(s_u)= 
 - \Big( \log \frac{\exp(sim(h_u, h^\mathrm{rtrl}_u)/\tau)}{\exp(sim(h_u, h^\mathrm{rtrl}_u)/\tau) + \sum_{s^- \in S_{u}^-} \exp(sim(h_u, h^-)/\tau)}+\\
&  \log \frac{\exp(sim(h^\mathrm{rtrl}_u,h_u)/\tau)}{\exp(sim(h^\mathrm{rtrl}_u,h_u)/\tau) + \sum_{s^- \in S_{u}^-} \exp(sim(h_u^\mathrm{rtrl}, h^-)/\tau)} \Big)
\end{split}
\end{equation*}
where $\tau$ is the temperature ratio.

\subsection{Experiment for Important Item Evaluation}

As shown in Figure~\ref{fig:example}, random augmentation is prone to generate false positive pairs that possibly degrade the quality of learned representations. To evaluate this claim, we conduct an experiment comparing CL4SRec using the vanilla random augmentation operations and augmentation operations that are aware of important items.  Our goal is to evaluate if the latter can contribute to better recommendation performance, suggesting that the item importance-aware approach generates higher quality representations.

To verify this assumption empirically, we construct a synthetic dataset\footnote{Details of the synthetic data is available at \url{https://github.com/demoleiwang/EC4SRec}.}, which provides ground truth of important items in every user sequence. 
Specifically, the dataset consists of $500$ user sequences each with 10 historical items and 3 additional items at the end serving as the next-items. Among the historical items are 3 important items shared by the 3 next-items to be used for training, validation and test respectively.  


We then experiment CL4SRec on this synthetic dataset with two types of mask operations to generate positive views.  Each of them masks $4$ historical items as follows: (i) \textit{random masking} that randomly masks $4$ historical items ($4$ is empirically chosen); and (ii) \textit{oracle based masking} that masks only unimportant items of the user sequence.

\begin{table}[t]
    \caption{Results of CL4SRec on synthetic dataset with ground truth important items.}
    \vspace{-10pt}
    \small
    \resizebox{0.26\textwidth}{!}{
    \begin{tabular}{l|c|c}
    \toprule
    Masking Op. & Random & Oracle-based \\ 
    \midrule
    HR@3 & 0.3560 & 0.5180 \\
    NDCG@3 & 0.2573 & 0.3645 \\
    \bottomrule
    \end{tabular}
    }
    \label{tab:toy}
    \vspace{-10pt}
\end{table}

As shown in Table~\ref{tab:toy}, CL4SRec using oracle-based masking substantially outperforms that using random masking by both HitRate@3 and NDCG@3. The former achieves more than 40\% higher NDCG@3 than the latter. This motivates us to determine  important items for effective augmentation and contrastive learning in sequential recommendation. 

\section{Explanation-Guided Contrastive Learning Approach}
\label{sec:method}



\begin{figure*}[t]
  \centering
  \includegraphics[width=0.96\linewidth]{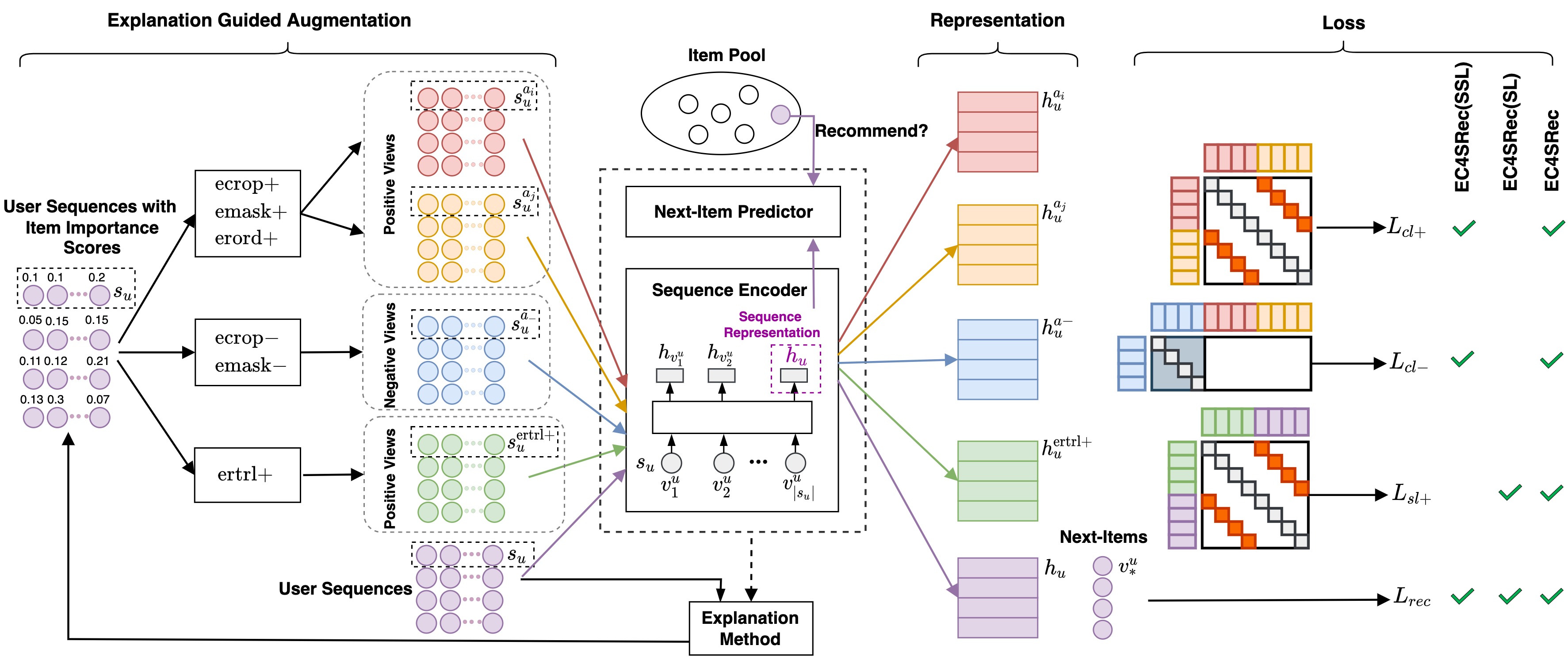}
  \vspace{-5pt}
  \caption{Proposed EC4SRec Framework.}
  \label{fig:overall}
\vspace{-10pt}
\end{figure*}

\subsection{Proposed Framework}
\label{sec:proposed_model}
Our proposed \textbf{\underline{E}xplanation guided \underline{C}ontrastive Learning Framework \underline{f}or  \underline{S}equential} \textbf{\underline{R}ecommendation} (\textbf{EC4SRec}), as shown in Figure~\ref{fig:overall}, consists of a \textbf{sequence encoder} to represent a given user's sequence of historical items $s_u$ into a vector representation $h_u$ which is in turn matched with items from a common pool by a \textbf{next-item predictor} which returns the next recommended item.

Unlike the existing contrastive learning methods to sequential recommendation (e.g., CL4SRec, DuoRec), EC4SRec utilizes an \textbf{explanation method} at scheduled epoch(es) to determine for a user sequence with next-item returned by the sequence encoder and next-item predictor the importance of each $s_u$'s items. Next, the \textbf{explanation guided augmentation} will utilize the item importance scores to generate positive and negative views of user sequences for further training the sequence encoder and next-item predictor.  The right of Figure~\ref{fig:overall} shows the different loss and recommendation loss functions that are used to train the models under different explanation-guided contrastive learning methods. 

The schedule of explanation method updating the item importance scores, also known as \textit{update schedule}, is controlled by a hyperparameter $p$. For a model training with a total of $N$ epoches, we schedule the updates to be at epoch $l  \cdot \lfloor \frac{N}{p+1} \rfloor$ for $1 \leq l \leq p$.  For example, for $p=3$ and $N=100$, updates will scheduled at epochs $25$, $50$, and $75$.  For epochs before the first scheduled update (i.e., 1 to $\lfloor \frac{N}{p+1} \rfloor -1$), EC4SRec can adopt any reasonably good sequential recommendation model (e.g., CL4SRec or DuoRec) to train the initial sequence encoder and next-item predictor.  In our experiments, we combine the losses of CL4SRec and DuoRec, i.e., $\sum_{u \in U}\mathcal{L}_{rec}(s_u)+ \lambda \mathcal{L}_{cl}(s_u)+ \lambda \mathcal{L}_{sl}(s_u)$, to train the initial model.
During inference, we only need to feed the input user sequence to the sequence encoder which generates the sequence representation for next-item predictor to recommend the next-item.

\subsection{Explanation Guided Importance Scores}

General explanation methods, such as Saliency Maps~\cite{zeiler2014visualizing}, Integrated Gradient~\cite{sundararajan2017axiomatic}, and Occlusion~\cite{simonyan2013deep}, are agnostic to sequential recommendation algorithms, 
such as GRU4Rec~\cite{tan2016improved}, Caser~\cite{tang2018personalized}, and SASRec~\cite{kang2018self}. 
To obtain explanation-guided importance scores for each item in the user sequence, we feed the input user sequence $s_u=[v^u_1, v^u_2, \cdots, v^u_{|s_u|}]$, the sequential encoder $SeqRec$, and its prediction probability for any next item $y_u$ into any model-agnostic explanation method $Expl(\cdot)$, which determine the \textbf{importance scores} of items in $s_u$ as $score(s_u) = Expl(y_u, s_u, SeqRec)$,
where $score(s_u) = [score(v^{u}_1), score(v^{u}_2),\dots, score(v^{u}_{|s_u|})]$ and $score(v^u_i)$ denotes the importance score of item $v^u_i$. 

While any explanation method (e.g., Saliency, Occlusion, and Integrated Gradient) can be used as $Expl(\cdot)$, we use Saliency to illustrate how importance score of each item is derived. Assume that there is an item embedding matrix $E\in \mathbb{R}^{|V|\times d}$, where $d$ is the embedding dimension. The embedding vector  $e_{v^{u}_i}\in\mathbb{R}^d$ of item $v^{u}_i$ can be derived from $E$. The importance score of dimension $j$ of $e_{v^{u}_i}$ can be defined by:
$
score(e_{{v}^{u}_{i,j}}) = \|\frac{\partial y_u}{\partial e_{v^{u}_{i,j}}}\|.
$

By adding and normalizing the importance scores of $d$ dimensions, we obtain the importance score of $e_{v^{u}_i}$, or $score(e_{v^{u}_i})$:
\begin{equation}
    score(v^{u}_i) = \frac{\sum_{j=1}^{d}score(e_{v^{u}_{i,j}})}{\sum_{i'=1}^{|s_u|}\sum_{j=1}^{d}score(e_{v^{u}_{i',j}})} .
\end{equation}

$score(v^u_i)$ returns a value in $[0,1]$ indicating how important is $v^u_i$ in the sequence $s_u$ for a specific given sequential recommendation algorithm. As $\sum_{i} score(v_i^u)=1$, the importance score is relative and comparable only among items of the same sequence.


\subsection{Explanation Guided Augmentation}

We propose five explanation guided augmentation operations, three for generating positive views and two for generating negative views. 
While some of these operations are extensions of random augmentation, the operations for generating negative views (that is, $\mathrm{ecrop-}$ and $\mathrm{emask-}$) are unique to EC4SRec as both CL4SRec and DuoRec consider augmentations for positive views only.  Our experiment results in Section~\ref{sec:expt_loss} also show that these negative views can substantially improve  recommendation accuracy.     
\begin{itemize}[leftmargin=*]
    
    \item \textbf{Explanation Guided Crop for Positive and Negative View ($\mathrm{ecrop+}$, $\mathrm{ecrop-}$):}
    To perform $\mathrm{ecrop+}$ (or $\mathrm{ecrop-}$) on $s_u$, we select the $k$ (or $|s_u|-k$) items with the lowest (or highest) by importance score to be removed to generate the positive (or negative) view. $k$ is defined by $\lfloor\mu_e \cdot |s_u|\rfloor$ where $\mu_e$ ($0 < \mu_e \leq 1$).
    Let $[v^u_{i_1}, \cdots, v^u_{i_k}]$ denote the sub-sequence of $k$ items in $s_u$ with the lowest importance scores.
    The explanation guided cropped positive and negative views are defined as:
    \begin{equation*}
    \begin{split}
      s_u^{\mathrm{ecrop+}}=s_u - [v^u_{i_1}, \cdots, v^u_{i_k}],         
    \end{split}
    \ \ \ \ \ \ 
    \begin{split}
      s_u^{\mathrm{ecrop-}}=[v^u_{i_1}, \cdots, v^u_{i_k}].      
    \end{split}
    \end{equation*}
    
    \item \textbf{Explanation Guided Mask for Positive or Negative View ($\mathrm{emask+}$,$\mathrm{emask-}$):}    
    To perform $\mathrm{emask+}$ on $s_u$, we select the $k$ items with the lowest importance scores to be masked. 
    Let $[v^u_{i_1}, \cdots\\, v^u_{i_k}]$ denote the sub-sequence of $k$ items in $s_u$ with the lowest importance scores.
    The explanation guided masked positive view is then defined as:
    \begin{equation*}
    \begin{split}
      s_u^{\mathrm{emask+}}=[v^u_1,\cdots,v^u_{i_1-1}, \mbox{[m]},v^u_{i_1+1}, \cdots,v^u_{i_k-1}, \mbox{[m]}, v^u_{i_k+1}, \cdots, v^u_{|s_u|}]
      \end{split}
    \end{equation*}
    The explanation guided masked negative view $s_u^{\mathrm{emask-}}$ is defined in a similar way except that the $k$ items to be masked are those with highest importance scores.
    
    \item \textbf{Explanation Guided Reorder for Positive View ($\mathrm{erord+}$):}
    Let $[v^u_{i_1}, \cdots, v^u_{i_k}]$ denote the sub-sequence of $k$ items in $s_u$ with the lowest importance scores ($i_1 < i_2 < \cdots < i_k$). We randomly reorder these items.  Suppose the reordered items, sorted by new positions, are $[\tilde{v}^u_{i_1}, \cdots, \tilde{v}^u_{i_k}]$. The reordered positive view can be formulated as:
    \begin{equation*}
      s_u^{\mathrm{erord+}}=[v^u_1,\cdots,v^u_{i_1-1}, \tilde{v}^u_{i_1}, v^u_{i_1+1}, \cdots, v^u_{i_k-1} \tilde{v}^u_{i_k},v^u_{i_k+1}, \cdots,v^u_{|s_u|}].
    \end{equation*}
    We leave out explanation guided reorder operation for negative views as it is not likely to generate discriminative negative views. 
    
    \item \textbf{Explanation Guided Retrieval for Positive View ($\mathrm{ertrl+}$):} Like random retrieval, we first define the candidate sequences that share the same target (or next) item as the original user sequence $s_u$ as: $S^c_u=\{s_{u_1},s_{u_2},\dots,s_{u_{|S^c_u|}}\}$.  That is, $v^{u_k}_*=v^u_*$, $s_{u_k}  \&\ u_k \neq u$. 
    Next, we compute the probability for each sequence in $S^c_u$ using importance scores:
    \begin{equation*}
    P(s_{u_k}) = \frac{util(s_{u_k})}{\sum_{s_{u_j}\in S^c_u} util(s_{u_j})}.
    \end{equation*}
    where 
    \begin{equation*}
     util(s_{u_k}) = \frac{|s_u\cap s_{u_k}|}{|s_u\cup s_{u_k}|} \sum_{v \in s_u\cap s_{u_k}} score(v)
    \end{equation*}
    
    We then sample the explanation guided retrieved sequence user sequence $s_u^{\mathrm{ertrl+}}$ from $S^c_u$ with the probability distribution. 
    
\end{itemize}


\subsection{Explanation Guided Contrastive Learning}
\label{sec:eg_CL}
Based on the EC4SRec framework, we can derive different proposed models depending the type of explanation guided contrastive learning used for model training.  In the following, we introduce three proposed models based on explanation guided self-supervised contrastive learning, explanation guided supervised contrastive learning, and combined explanation guided contrastive learning.  


\subsubsection{Explanation Guided Self-Supervised Learning (\textbf{EC4SRec(SSL)})}

This model can be seen as an extension of CL4SRec with explanation guided augmentation operations generating both positive and negative views for contrastive learning. The loss function consists of three components: (i) \textit{recommendation loss}, (ii) \textit{contrastive loss for explanation guided positive views}, and (iii) \textit{contrastive loss for explanation guided negative views}:
\begin{equation}
\begin{split}
\mathcal{L}_{EC4SRec(SSL)} = \sum_{u \in U_B} \mathcal{L}_{rec}(s_u) + 
 \lambda_{cl+}  (\mathcal{L}_{cl+}(s_u) +     
 \lambda_{cl-} \mathcal{L}_{cl-}(s_u)). 
\end{split}
\end{equation}

The $\mathcal{L}_{rec}(s_u)$ here has been defined earlier in Equation 2.  
Let $A^+ =\{ a_\mathrm{ecrop+}, a_\mathrm{emask+}, a_\mathrm{erord+}\}$ and $A^- =\{ a_\mathrm{ecrop-}, a_\mathrm{emask-}\}$. To obtain $\mathcal{L}_{cl+}(s_u)$, we generate two positive views $s_u^{a_i}$ and $s_u^{a_j}$ by sampling $a_i$ and $a_j$ ($a_i \neq a_j$) from $A^+$ and applying them on $s_u$. We repeat this for all other users and obtain a set of a set of positive views from \textit{all} users denoted as $S^{+}$.  Let $S^+_u$ be $\{s_u^{a_i},s_u^{a_j}\}$. 
To get the representations of $s_u^{a_i}$ and $s_u^{a_j}$ closer to each other but farther away from other users' positive views, we define: 
\begin{equation*}
\begin{split}
\mathcal{L}_{cl+}(s_u)=
 -\log \frac{\exp(sim(h_u^{a_i}, h_u^{a_j}))}{\exp(sim(h_u^{a_i}, h_u^{a_j}))+\sum_{s^a_{u'} \in S^+ - S^+_u} \exp(sim(h_u^{a_i}, h^a_{u'}))}
\end{split}
\end{equation*}
To obtain $\mathcal{L}_{cl-}(s_u)$, we generate a negative view $s_u^{a-}$ by applying an augmentation operator $a-$, sampled from $A^-$, on $s_u$. Here, we would like this negative view to be closer to other users' negative views (as we do not need distinctive representations for these negative views) and farther away from the all users' positive views, borrowing a similar idea from ~\cite{khosla2020supervised}.  
Let $S^{-}$ denote the set of negative views after repeating the above on \textit{all} users. 
We define:
\begin{equation*}
\begin{split}
\mathcal{L}_{cl-}(s_u)=
 -\frac{1}{|S^{-}|-1}\sum_{s_{u'}^{a} \in S^{-}-\{s^{a-}_u\}}\log \frac{\exp(sim(h_u^{a-}, h_{u'}^{a}))}{\sum_{s \in S^+ \cup \{s^{a}_{u'}\}} \exp(sim(h_u^{a-}, h))},
\end{split}
\end{equation*}
where $h$ is the representation of the sequence $s$. By setting $\beta=0$, we can obtain a model variant that considers positive views only.


\subsubsection{Explanation Guided Supervised Contrastive Learning\\ (\textbf{EC4SRec(SL)})}

This model extends DuoRec to use explanation guided augmentation.
The loss function is:
\begin{equation}
\mathcal{L}_{EC4SRec(SL)} = \sum_{u \in U_B} \mathcal{L}_{rec}(s_u) +  \lambda  \mathcal{L}_{sl+}(s_u), 
\end{equation}
where 
\begin{equation}
\begin{split}
\mathcal{L}_{sl+}&(s_u)= \\
& - \Big( \log \frac{\exp(sim(h_u, h^{ertrl+}_u)/\tau)}{\exp(sim(h_u, h^{ertrl+}_u)/\tau) + \sum_{s^- \in S_{u}^-} \exp(sim(h_u, h^-)/\tau)}+\\
&  \log \frac{\exp(sim(h^{ertrl+}_u,h_u)/\tau)}{\exp(sim(h^{ertrl+}_u,h_u)/\tau) + \sum_{s^- \in S_{u}^-} \exp(sim(h_u^{ertrl+}, h^-)/\tau)} \Big).
\end{split}
\end{equation}

$h_u^{ertrl+}$ is the representation of the augmented sequence for the user $u$ generated by explanation guided retrieval operation, i.e., $\mathrm{ertrl+}$. 

\subsubsection{Combined Explanation Guided Contrastive Learning (\textbf{EC4SRec})}
To leverage both self-supervised contrastive learning and supervised contrastive learning, we combine two contrastive learning losses as:
\begin{equation}
\begin{split}
\mathcal{L}&_{EC4SRec} = \\
& \sum_{u \in U_B} \mathcal{L}_{rec}(s_u) + 
 \lambda_{cl+} \mathcal{L}_{cl+}(s_u) +     
 \lambda_{cl-} \mathcal{L}_{cl-}(s_u) +
 \lambda_{sl+}
 \mathcal{L}_{sl+}(s_u)
\end{split}
\label{equ:27}
\end{equation}

\section{Experiment}

\subsection{Experimental Settings}
\subsubsection{Datasets and Data Preprocessing}

\begin{table}[t]
    \caption{Dataset Statistics After Preprocessing.}
    \vspace{-10pt}
    \small
    \resizebox{0.4\textwidth}{!}{
    \begin{tabular}{l|rrrr}
    \toprule
    Dataset & Beauty  & Clothing & Sports & ML-1M  \\
    \midrule
    Users & 22,363 & 39,387 & 35,598 & 6,041 \\
    Items & 12,101 & 23,033 & 18,357 & 3,417 \\
    User-item Interactions & 198,502 & 278,677 & 296,337 & 999,611 \\
    Avg Sequence Length & 8.9 & 7.1 & 8.3 & 165.5 \\
    Sparsity & 99.93\% & 99.97\% & 99.95\% & 95.16\%\\
    \bottomrule
    \end{tabular}
    }
    \label{tab:datasets}
    \vspace{-10pt}
\end{table}

\begin{table*}[t]
    \caption{Overall Results. 
(The best and second best results are boldfaced and underlined. *: significant improvement of EC4SRec(SSL) over CL4SRec with $p$-value$=0.05$. **:  significant improvement of EC4SRec(SL) over DuoRec with $p$-value$=0.01$.)}
    \vspace{-10pt}
    \small
    \resizebox{0.94\textwidth}{!}{
    \begin{tabular}{c|l|c|cccc|c|cg|cg|g}
    \toprule
         &  & Non-Seq. & \multicolumn{5}{c|}{Seq. Rec. w/o Contrastive Learning} & \multicolumn{5}{c}{Seq. Rec. with Contrastive Learning} \\
         \hline
         Dataset& Metric & BPR-MF & GRU4Rec & Caser & SASRec & BERT4Rec &$\text{S}^3\text{Rec}_\text{MIP}$ & CL4SRec & EC4SRec(SSL)** & DuoRec & EC4SRec(SL)* & \textbf{EC4SRec} \\
         \midrule
         \multirow{4}*{Beauty}&HR@5&0.0120&0.0164&0.0191&0.0365&0.0193&0.0327&
         0.0495&0.0569&
          
         0.0548&  \bf 0.0585 & \underline{0.0569} \\
         &HR@10&0.0299&0.0365&0.0335&0.0627&0.0401&0.0591&
         0.0810&0.0853&
          
         0.0832&  \bf 0.0867 & \underline{0.0862}\\
         &NDCG@5&0.0065&0.0086&0.0114&0.0236&0.0187&0.0175&
         0.0299&0.0358&
          
         0.0345&  \underline{0.0361} &\bf  0.0364 \\
         &NDCG@10&0.0122&0.0142&0.0160&0.0281&0.0254&0.0268&
         0.0401&  0.0450&
          
         0.0436& \underline{0.0455} & \bf 0.0458 \\
         \midrule
         \multirow{4}*{Clothing}&HR@5&0.0067&0.0095&0.0049&0.0168&0.0125&0.0163&
         0.0187&0.0201&
          
         0.0196& \underline{0.0205} & \bf0.0209\\
         &HR@10&0.0094&0.0165&0.0092&0.0272&0.0208&0.0237&
         0.0305&\underline{0.0314}&
          
         0.0296& 0.0311 &\bf 0.0320\\
         &NDCG@5&0.0052&0.0061&0.0029&0.0091&0.0075&0.0101&
         0.0104&0.0113&
          
         0.0112& \underline{0.0115} & \bf0.0119\\
         &NDCG@10&0.0069&0.0083&0.0043&0.0124&0.0102&0.0132&
         0.0142&\underline{0.0149}&
          
         0.0144& \underline{0.0149} & \bf0.0155\\
         \midrule
         \multirow{4}*{Sports}&HR@5&0.0092&0.0137&0.0121&0.0218&0.0176&0.0157&
         0.0277&\underline{0.0323}&
          
         0.0310& 0.0317 & \bf0.0331 \\
         &HR@10&0.0188&0.0274&0.0204&0.0336&0.0326&0.0265&
         0.0455&\underline{0.0497}&
          
         0.0480& 0.0491 &\bf 0.0514\\
         &NDCG@5&0.0053&0.0096&0.0076&0.0127&0.0105&0.0098&
         0.0167&\underline{0.0201}&
          
         0.0191& 0.0194& \bf 0.0203\\
         &NDCG@10&0.0085&0.0137&0.0103&0.0169&0.0153&0.0135&
         0.0224&\underline{0.0256}&
          
         0.0246& 0.0249 & \bf 0.0262\\
         \midrule
         \multirow{4}*{ML-1M}&HR@5&0.0164&0.0763&0.0816&0.1087&0.0733&0.1078&
         0.1583&\bf0.1699&
          
         0.1672&\underline{0.1682} & 0.1672\\
         &HR@10&0.0354&0.1658&0.1593&0.1904&0.1323&0.1952&
         0.2423&\bf0.2543&
          
         0.2507&0.2526 & \underline{0.2533}\\
         &NDCG@5&0.0097&0.0385&0.0372&0.0638&0.0432&0.0616&
         0.0996&0.1095&
          
         0.1076&\bf0.1104 & \underline{0.1102}\\
         &NDCG@10&0.0158&0.0671&0.0624&0.0910&0.0619&0.0917&
         0.1267&0.1368&
          
         0.1345&\underline{0.1375} & \bf0.1380\\
         \bottomrule
    \end{tabular}
    }
    \label{tab:overall}
    \vspace{-5pt}
\end{table*}

We conduct experiments on four widely used real-world datasets, i.e., Beauty, Clothing, Sports, and ML-1M (Movielens 1M). The first three are from Amazon\footnote{http://jmcauley.ucsd.edu/data/amazon/}~\cite{mcauley2015image}, and ML-1M\footnote{https://grouplens.org/datasets/movielens/1m/}~\cite{harper2015movielens} is a very large benchmark dataset for movie recommendation.
Following previous works ~\cite{kang2018self, zhou2020s3, xie2020contrastive, qiu2022contrastive},  we  remove repeated items, and preprocess four datasets with the 5-core strategy (i.e., removing users and items with fewer than 5 interactions).
The dataset statistics are summarized in Table~\ref{tab:datasets}.  The datasets are very sparse as their sparsity indices (defined by $1-\frac{\mbox{num. of interactions}}{\mbox{num. of users} \cdot \mbox{num. of items}}$) are very high.


\subsubsection{Evaluation Protocols}
Following~\cite{xie2020contrastive, qiu2022contrastive}, 
we use the last interacted item of each user sequence for test, the second last item for validation, and all the earlier items for training. 
The predicted next-item come from the pool of \textit{all} items without any candidate filter. We employ two performance metrics, \textbf{Hit Ratio at $k$ (HR@k)} and \textbf{Normalized Discounted Cumulative Gain at $k$ (NDCG@k)}, which are widely used in previous work~\cite{kang2018self, zhou2020s3, xie2020contrastive, qiu2022contrastive}. 
We report the average of results with running 3 times with 3 random seeds.

\subsubsection{Baselines}

\begin{table*}
\caption{Results of EC4SRec with different Sequential Recommendation Backbones.}
\vspace{-10pt}
\setlength{\tabcolsep}{0.35em}
\centering
\small
\resizebox{0.92\textwidth}{!}{

\begin{tabular}{l|l|cccc|cccc|cccc }
\toprule

 &  &\multicolumn{4}{c|}{Beauty}&\multicolumn{4}{c|}{Clothing}&\multicolumn{4}{c}{Sports}\\
\rowcolor{white} Backbone &  & HR@5 & HR@10 & NDCG@5 & NDCG@10 & HR@5 & HR@10 & NDCG@5 & NDCG@10 & HR@5 & HR@10 & NDCG@5 & NDCG@10 \\ \midrule
\multirow{6}{*}{GRU4Rec}& CL4SRec & 0.0420 & 0.0640 & 0.0270 & 0.0341 & 0.0104 & 0.0180 & 0.0065 & 0.0089 & 0.0244 & 0.0389 & 0.0154 & 0.0200    \\
 &  EC4SRec(SSL)&0.0461 & 0.0674 & 0.0314 & 0.0382 & 0.0128 & \underline{0.0213} & 0.0082 & \underline{0.0109} & 0.0253 & 0.0396 & 0.0167 & 0.0213  \\ \cline{ 2-14}
& DuoRec & 0.0471 & 0.0689 & 0.0318  & 0.0388 &  0.0118 & 0.0193 & 0.0078 & 0.0102 & 0.0259 & 0.0396 & 0.0163 & 0.0207   \\
&   EC4SRec(SL) &  \underline{0.0490} & \underline{0.0717} & \underline{0.0327} & \underline{0.0401} & \underline{0.0130} & 0.0201 & \underline{0.0086} & 0.0108 &  \underline{0.0273} & \underline{0.0414} & \underline{0.0173} & \underline{0.0218}   \\ \cmidrule{ 2-14}
 &  EC4SRec & \bf 0.0495 &\bf0.0745 & \bf0.0332 & \bf0.0412 & \bf0.0138 & \bf0.0218 & \bf0.0089 &\bf 0.0115 & \bf 0.0276 & \bf0.0437 &\bf 0.0182 & \bf0.0233   \\  
\midrule

\multirow{6}{*}{Caser} & CL4SRec & 0.0185 & 0.0335 & 0.0108 & 0.0157 & 0.0058 & 0.0100 & 0.0036 & 0.0049 & 0.0113 & 0.0191 & 0.0071 & 0.0096   \\
&   EC4SRec(SSL) &0.0228 &0.0390 & 0.0137 & 0.0189 & \underline{0.0064} & 0.0113 & \underline{0.0039} & 0.0055 & 0.014 & \underline{0.0244} & 0.0088 & 0.0121   \\ \cline{ 2-14}
& DuoRec & 0.0207 & 0.0375 & 0.0129 & 0.0183 & 0.0053 & 0.0100 & 0.0031 & 0.0046 & 0.0127 & 0.0215 & 0.0082 & 0.0110   \\
&  EC4SRec(SL) & \underline{0.0262} & \underline{0.0439} & \underline{0.0161} & \underline{0.0218} & \underline{0.0064} & \underline{0.0117} & \underline{0.0039} & \underline{0.0056} & \underline{0.0146} & 0.0240 & \underline{0.0097} & \underline{0.0127}   \\ \cline{ 2-14}

&  EC4SRec & \bf0.0269 & \bf0.0456 & \bf0.0172 & \bf0.0232 & \bf 0.0065 & \bf 0.0124 &\bf 0.0041 & \bf0.0060 &  \bf0.0152 & \bf0.0266 & \bf0.0105 & \bf0.0139  \\

\bottomrule
\end{tabular}
}

\label{tab:plug_and_play}

\vspace{-10pt}
\end{table*}

We compare EC4SRec(SSL), EC4SRec(SL) and EC4SRec with the following three groups of baseline methods:
\begin{itemize}[leftmargin=*]
    \item \textbf{Non-sequential recommendation method}: We use BRP-MF~\cite{rendle2012bpr}, a popular matrix factorization model. 
    \item \textbf{Sequential recommendation methods without contrastive learning:} GRU4rec~\cite{hidasi2015session}, a RNN-based method; Caser~\cite{tang2018personalized}, a CNN-based method; two self-attention based methods SASRec~\cite{kang2018self} and BERT4Rec~\cite{sun2019bert4rec}; and a self-supervised learning method  S$^3$Rec$_{\mathrm{MIP}}$ \cite{zhou2020s3}.
    \item \textbf{Sequential recommendation methods with contrasitve learning}:  CL4SRec~\cite{xie2020contrastive} and DuoRec~\cite{qiu2022contrastive}. 
\end{itemize}

\subsubsection{Implementation Details}

For BPR-MF and S$^3$Rec$_{\mathrm{MIP}}$, we use results reported by CL4SRec~\cite{xie2020contrastive}.  We implemented the baselines GRU4Rec, Caser, SASRec, and BERT4Rec using the RecBole library
\footnote{\url{https://github.com/RUCAIBox/RecBole}}
\cite{zhao2021recbole}. For CL4SRec and DuoRec, we made some changes to the codes
provided by the authors of DuoRec to mainly correct some bugs. 
Our CL4SRec and DuoRec results are generally  similar to that reported in the original works. For each baseline, we set the embedding dimension to be $64$ and keep all other hyper-parameter settings the same as those reported in their original papers.  
For EC4SRec and its variants, we use SASRec and Occlusion as the default backbone sequential recommendation method and explanation method respectively.  
The hyper-parameter settings, such as batch size, embedding dimension, number of layers, number of attention heads, follow those reported in \cite{xie2020contrastive}. We tune $\mu_e$, a hyperparameter to control the proportion of important items in augmentation from $0.1$ to $0.9$ with step size $=0.1$.  We also tune the temperature $\tau$ within $[0.5, 1.0, 1.5]$, and the coefficients $\lambda_{cl+}$, $\lambda_{cl-}$, and $\lambda_{sl+}$ within $[0.1, 0.2, 0.3, 0.4, 0.5]$. 

\subsection{Overall Results}

\subsubsection{EC4SRec versus Baselines.}
\label{sec:ec4srec_vs_baselines}
As shown in Table~\ref{tab:overall}, the overall experiment results show that:

\begin{itemize}[leftmargin=*]

    \item Our proposed EC4SRec and its variants consistently outperform the state-of-the-art methods, including the latest contrastive learning-based models CL4SRec 
    and DuoRec,
    for all datasets by all metrics. Specifically, EC4SRec achieves 12.4\% (4.9\%) improvement over CL4SRec (DuoRec) on average across all datasets by all metrics. EC4SRec generally performs better than EC4SRec(SSL) and EC4SRec(SL).  The above findings as well as the significant improvement of EC4SRec(SSL) over CL4SRec and EC4SRec(SL) over DuoRec demonstrate that  self-supervised and supervised contrastive learning benefit substantially from explanation guided augmentation.  Higher-quality positive views and negative views for contrastive leaning have clearly resulted in better user sequence representations. 
    
    \item Among the baselines, non-sequential recommendation recommendation methods (i.e., BPR-MF) unexpectedly yield the worst performance. It indicates that the sequential patterns are important in this task. 
    Among the sequential recommendation methods, SASRec and BERT4Rec consistently outperform 
    GRU4Rec and Caser.
    It shows that self attention can model more complex patterns than left-to-right patterns.
    
    \item Consistent with earlier results in \cite{xie2020contrastive,qiu2022contrastive}, contrastive learning methods CL4SRec and DuoRec outperform S$^3$Rec$_{\mathrm{MIP}}$ and SASRec. Our experiment also shows that the former also outperform BERT4Rec. The above demonstrates the the strength of contrastive learning.  With supervised contrastive learning, DuoRec performs better than CL4SRec but the gap is reduced between EC4SRec(SL) and EC4SRec(SSL). This could be explained by the additional loss $\mathcal{L}_{cl-}$ added to EC4SRec(SSL).  


\end{itemize}

\subsubsection{EC4SRec with Different Backbone Sequential Recommendation Methods}

Instead of using the default self-attention based backbone SASRec, we evaluate EC4SRec, its variants, CL4SRec and DuoRec using other backbones, namely RNN-based GRU4Rec and CNN-based Caser to study the impact of explanation guided augmentation and contrastive learning. Due to space constraint, we only show the result on three datasets. 
As shown in Table~\ref{tab:plug_and_play}, the relative performance ranking between EC4SRec, EC4SRec(SSL), EC4SRec(SL), CL4SRec and DuoRec remains unchanged when using GRU4Rec and Caser backbones. EC4SRec still achieves the best performance using different backbones. EC4SRec(SSL) and EC4SRec(SL) outperforms CL4SRec and DuoRec respectively. These encouraging results indicate the generalizability of the effectiveness of explanation guided approach. 



\subsection{Detailed Analysis}

In this section, we conduct detailed analysis of EC4SRec and its variants. 
We show the results on Beauty and Clothing datasets only due to space constraint. 

\subsubsection{Effect of Update Schedule of Important Scores}

    

\begin{figure}[t]
    \centering
    
    \includegraphics[width=1.0\linewidth]{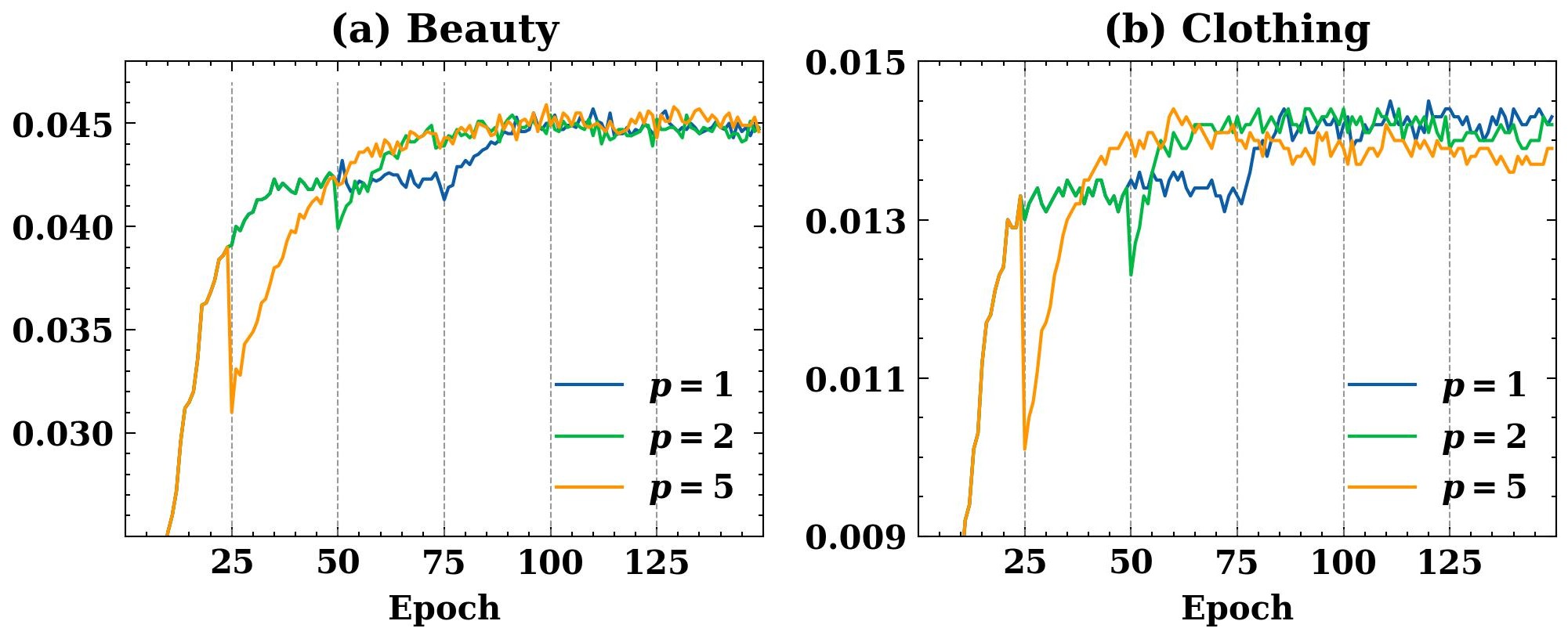}
    \vspace{-15pt}
    \caption{Changes of NDCG@5 for EC4SRec using different update schedules over 150 training epoches ($p$: number of importance score updates in training)}
    \label{fig:train_process}
    \vspace{-10pt}
\end{figure}

\begin{figure}[t]
    \centering
    
    \includegraphics[width=1.0\linewidth]{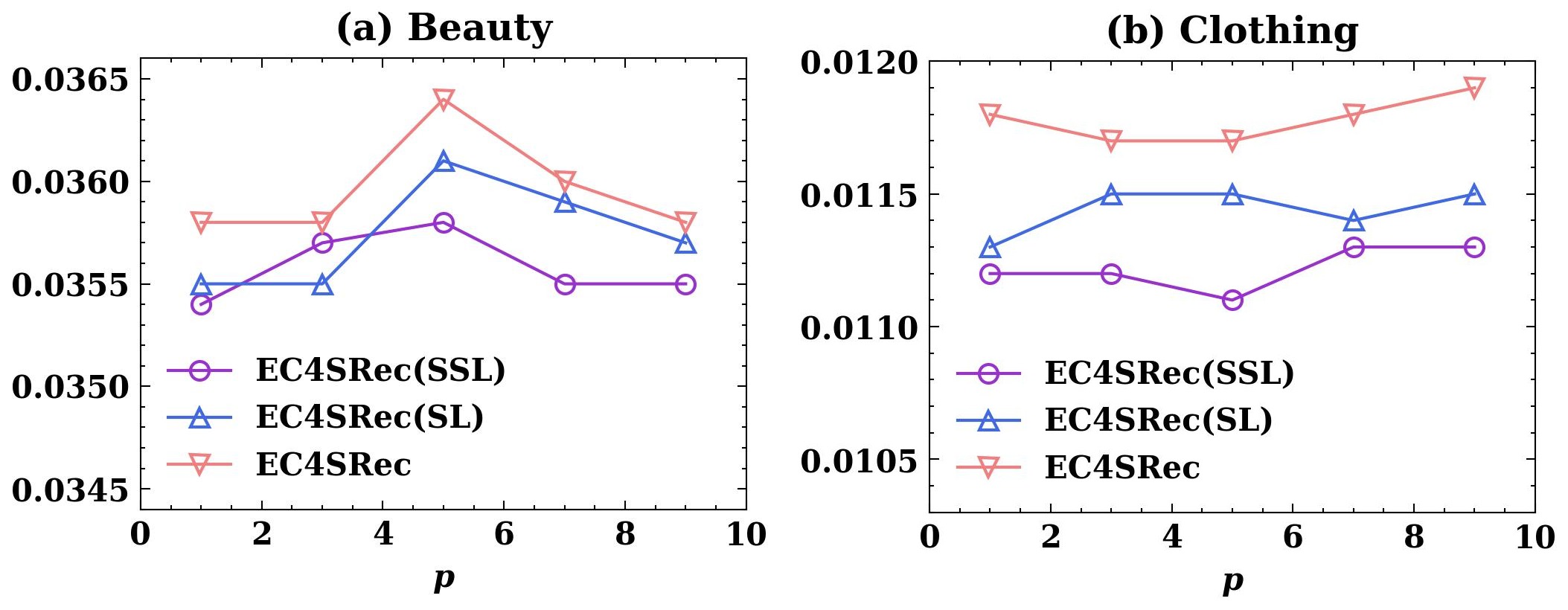}
    \vspace{-15pt}
    \caption{NDCG@5 Results with different  update schedule settings ($p$: number of updates in model training). 
    }
    \label{fig:freq}
    \vspace{-10pt}
\end{figure}

As mentioned in Section~\ref{sec:proposed_model}, the parameter $p$ controls the number of item importance updates scheduled during the  training of EC4SRec and its variants.  First, we want to study how the updates affect the performance of these models during the training epoches.  For illustration, we plot the NDCG@5 of EC4SRec only on validation data in Figure~\ref{fig:train_process}. With 150 training epoches, the update occurs only  at epoch 75 for $p=1$, at epoches 50 and 100 for $p=2$, and at epoches 25, 50, 75, 100 and 125 for $p=5$.  The figure shows that EC4SRec experiences drops of NDCG@5 at the first update.
This is because EC4SRec switches from random augmentation and a loss function combining that of CL4SRec and DuoRec to explanation guided augmentation and explanation guided contrastive loss at the first update. EC4SRec however recovers quickly and continues to improve until it converges. Interestingly, the drop in performance is not noticeable for subsequent scheduled updates. 

We also show the effect of update schedule on the trained EC4SRec and variants when evaluated against test data in Figure~\ref{fig:freq}.  Generally, the NDCG@5 performance does not change much for different $p$ settings.  $p=5$ and $=9$ yield best results for Beauty and Clothing respectively. 
As every update incurs additional overheads, there is clearly a trade-off between performance and efficiency when choosing the update schedule which we shall leave to future research.

\subsubsection{Ablation of Loss Functions.}
\label{sec:expt_loss}

\begin{figure}[t]
    \centering
    
    \includegraphics[width=0.92\linewidth]{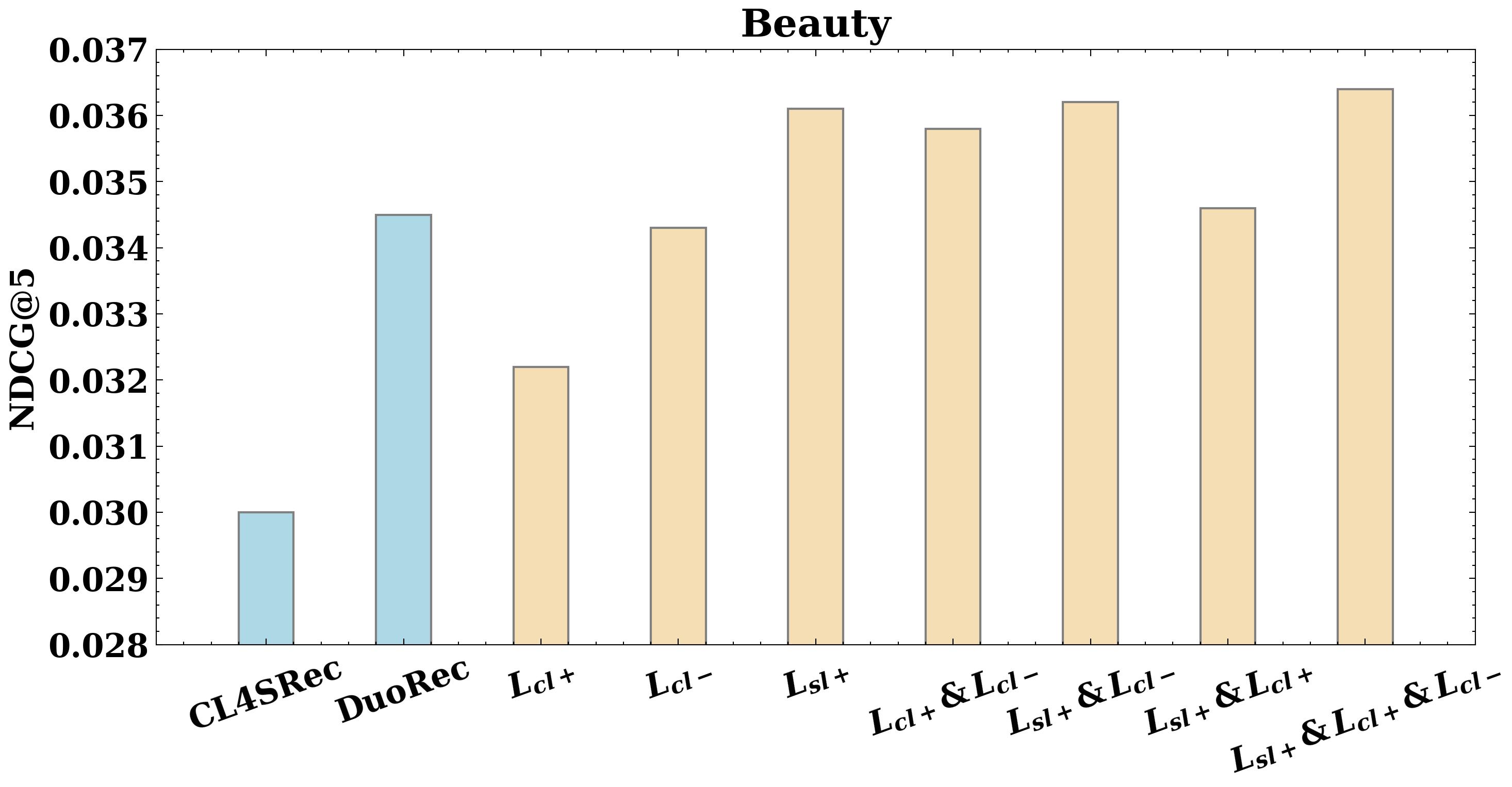}
    \vspace{-10pt}
    \caption{Ablation study of EC4SRec with different combinations of loss functions on Beauty dataset. (EC4SRec results are shown in yellow bars.  As $\mathcal{L}_{rec}$ is included by default, EC4SRec(SSL) = EC4SRec with $\mathcal{L}_{cl+}+\mathcal{L}_{cl-}$; EC4SRec(SL) = EC4SRec with $\mathcal{L}_{sl+}$, and EC4SRec = one with all three losses.) 
    }
    \label{fig:ablation}
    \vspace{-10pt}
\end{figure}

We study the effect of different contrastive losses on EC4SRec performance by evaluating the model using different combinations of losses as shown in Figure~\ref{fig:ablation}.  The figure shows the NDCG@5 of EC4SRec using recommendation loss $\mathcal{L}_{rec}$ and seven combinations of the three contrastive losses, $\mathcal{L}_{cl+}$, $\mathcal{L}_{cl-}$ and $\mathcal{L}_{sl+}$.  For illustration, we conduct this ablation study on Beauty and include CL4SRec and DuoRec for comparison.  

Figure~\ref{fig:ablation} shows that EC4SRec with $\mathcal{L}_{cl+}$ and EC4SRec with $\mathcal{L}_{cl-}$ outperform CL4SRec. EC4SRec with  $\mathcal{L}_{sl+}$ also outperforms DuoRec. These indicate that each of the 3 explanation guided contrastive losses can effectively improve performance. Moreover, combining them together can yield even higher performance with the exception of $\mathcal{L}_{sl+} + \mathcal{L}_{cl+}$ which could be explained by having only $\mathcal{L}_{cl+}$ without $\mathcal{L}_{cl-}$ does not help to improve representations much and may conflict with the supervised contrastive learning loss using retrieved positive views.


\subsubsection{Influence of Different Augmentation}

\begin{table}[t]
    \caption{NDCG@5 Results of EC4SRec(SSL), abbreviated by E(SSL), with the removal of augmentation operation on Beauty, Clothing and Sports.}
    \vspace{-10pt}
    \centering
    \resizebox{0.46\textwidth}{!}{
    \small
    \begin{tabular}{l|cg|cg|cg}
	    \toprule
	    
	     & \multicolumn{2}{c|}{Beauty} & \multicolumn{2}{c|}{Clothing}& \multicolumn{2}{c}{Sports} \\
	      & CL4SRec & E(SSL) &CL4SRec & E(SSL)&CL4SRec & E(SSL) \\
	    \midrule
	    None & 0.0299 &  0.0358  & 0.0104 &  0.0113 & 0.0167 & 0.0201 \\
	    \,\,$-$rord     & 0.0307 &  0.0344  &0.0103 &  0.0110 &0.0169&0.0181 \\
	    \,\,$-$mask &0.0311 &  0.0350  & 0.0101 & 0.0116 &0.0169&0.0200\\
	    \,\,$-$crop & 0.0282 &  0.0353  & 0.0086 &  0.0116 &0.0147&0.0200 \\
        \bottomrule
    \end{tabular}
    }
    
    \label{tab:aug}
    \vspace{-5pt}
\end{table}

Our approach consists of four explanation guided augmentations: ecrop, emask, erord, and ertrl. We earlier show that EC4SRec(SL) using explanation guided retrieval (i.e., ertrl) significantly outperforms DuoRec as shown in Table~\ref{tab:overall}. In this section, we evaluate how EC4SRec performs when not using one of the three augmentation operations to investigate the effect of each augmentation operation.  

As shown in Table~\ref{tab:aug}, the recommendation accuracy drops substantially when any one of augmentations is removed. Besides, EC4SRec(SSL) consistently achieves better performance than CL4SRec even with one augmentation operation removed. It indicates the effectiveness of each proposed operation. Interestingly, for Clothing dataset, the removal of some augmentation operation can slightly improve the EC4SRec(SSL) performance.


\subsubsection{Study of Hyper-Parameter $\mu_e$}


\begin{figure}[t]
    \centering
    
    \includegraphics[width=1.0\linewidth]{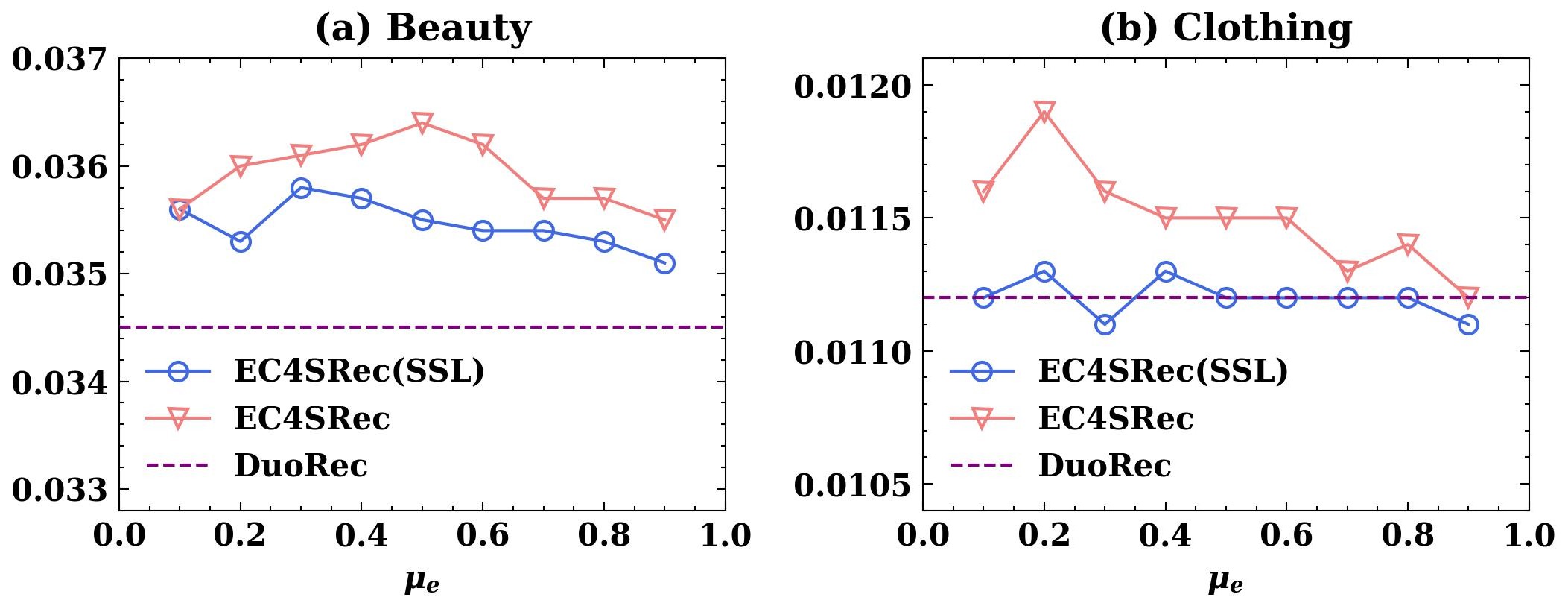}
    \vspace{-15pt}
    \caption{NDCG@5 of EC4SRec with different $\mu_e$ settings.}
    \label{fig:mu}
    \vspace{-10pt}
\end{figure}


The hyper-parameter $\mu_e$ determines the number of items with highest scores would be augmented for positive views and negative views under explanation guided augmentation.  In this study, we vary  $\mu_e$ from 0.1 to 0.9 and show the NDCG@5 of EC4SRec and EC4SRec(SSL) on Beauty and Clothing datasets as shown in Figure~\ref{fig:mu}. We observe that $\mu_e$ substantially affects the performance of EC4SRec and EC4SRec(SSL). 
For Beauty dataset, the NDCG@5 of EC4SRec changes from $0.0364$ when $\mu_e=0.5$ to $0.0355$ when $\mu_e=0.9$. 
Second, EC4SRec and EC4SRec(SSL) perform best on Beauty when $\mu_e=0.5$ and $\mu_e=0.3$ respectively.  
For Clothing dataset, the NDCG@5 of EC4SRec changes from $0.0119$ when $\mu_e=0.2$ to $0.0112$ when $\mu_e=0.9$. And both EC4SRec and EC4SRec(SSL) perform best on Clothing when $\mu_e=0.2$. These findings indicate that EC4Srec and its variants have different optimal value $\mu_e$ on different datasets. Besides, even EC4SRec with the worst performing $\mu_e$ still outperforms DuoRec.

\subsubsection{Effect of Different Explanation Methods}



\begin{figure}[t]
    \centering
    
    \includegraphics[width=1.0\linewidth]{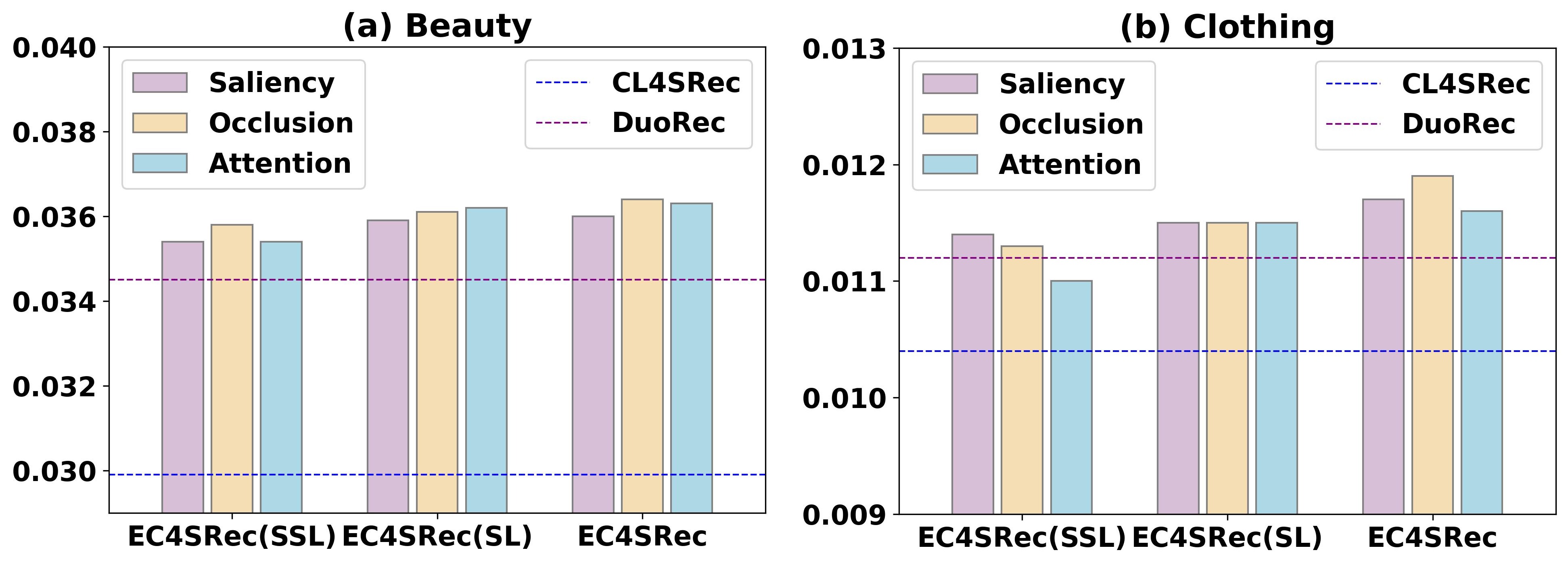}
    \vspace{-15pt}
    \caption{NDCG@5 using different explanation methods.}
    \label{fig:diff_x}
    \vspace{-10pt}
\end{figure}

In our earlier results, we use Occlusion as the default explanation method.  In this experiment, we evaluate EC4SRec and its variants using other explanation methods for comparison. Figure~\ref{fig:diff_x} shows the NDCG@5 results of the above models using Saliency, Occlusion, and Attention based explanation methods on Beauty and Clothing datasets.  The NDCG@5 of CL4SRec and DuoRec are also shown as reference. 

Figure~\ref{fig:diff_x} shows that Occlusion performs well in most cases. The three explanation methods generally help EC4SRec and variants outperform CL4SRec and DuoRec except Attention which could not help EC4SRec(SSL) outperforms DuoRec. Considering robustness, performance, and efficiency, we prefer to use occlusion as the explanation method to get better views for contrastive learning.

\section{Conclusion}

In this paper, we study how to utilize explanation methods to produce high-quality views for contrastive learning in sequential recommendation task. We propose a model-agnostic Explanation Guided Contrastive Learning for Sequential Recommendation (EC4SRec) framework. We introduce several proposed explanation-guided augmentations to generate positive and negative views of given user sequences and propose both self-supervised and supervised contrastive learning.  Our extensive experiments on four real-world benchmark datasets demonstrate the effectiveness, generality, and flexibility of our proposed explanation guided approach. Our results also outperform the state-of-the-art contrastive learning based models.  To our knowledge, this work represents the first that combine sequential recommendation with explanation methods. For our future research, we will conduct more extensive analysis of the importance score functions and training efficiency. Explanation guided supervised contrastive learning in particular could be slow as it involves selection of retrieved positive views using importance score function.  One future research direction is thus to address such overheads by developing appropriate indexing or hashing techniques. One another direction is to meet the challenge of designing augmentations for very long sequential recommendations in contrastive learning.

\section*{Acknowledgement} 
This research is supported by the National Research Foundation,
Singapore under its Strategic Capabilities Research Centres Funding Initiative. Any opinions, findings and conclusions or recommendations expressed in this material are those of the author(s) and do not reflect the views of National Research Foundation, Singapore.

\bibliographystyle{ACM-Reference-Format}
\bibliography{sample-base}

\end{document}